%% file: main.tex
\documentclass[a4paper, amsfonts, amssymb, amsmath, reprint, showkeys, nofootinbib, twoside,floatfix,aip]{revtex4-2}

\usepackage[english]{babel}
\usepackage[utf8]{inputenc}
\usepackage[colorinlistoftodos, color=green!40, prependcaption]{todonotes}
\input{preamble}

\usepackage[caption=false]{subfig}
\usepackage[pdftex, pdftitle={Article}, pdfauthor={Author}]{hyperref} 
\begin{document}

\title{Interplay of electron and phonon channels in the refrigeration through molecular junctions}

\author{Fatemeh Tabatabaei}

\author{Samy Merabia}
 \affiliation{Université Claude Bernard Lyon 1, CNRS, Institut Lumière Matière, Villeurbanne, France}
\author{Bernd Gotsmann}
    \affiliation{IBM Research Europe - Zurich, Rueschlikon, Switzerland}
\author{Mika Prunnila}
    \affiliation{VTT Technical Research Centre of Finland Ltd., Tietotie 3, FI-02150 Espoo, Finland}
\author{Thomas A. Niehaus}
\affiliation{Université Claude Bernard Lyon 1, CNRS, Institut Lumière Matière, Villeurbanne, France}
\email[Email address:]{thomas.niehaus@univ-lyon1.fr}
\date{\today} 

\begin{abstract}
Due to their structured density of states, molecular junctions provide rich resources to filter and control the flow of electrons and phonons. Here we compute the out of equilibrium current-voltage characteristics and dissipated heat of some recently synthesized oligophenylenes (OPE3) using the Density Functional based Tight-Binding (DFTB) method within Non-Equilibrium Green's Function Theory (NEGF). We analyze the Peltier cooling power for these molecular junctions as function of a bias voltage and investigate the parameters that lead to optimal cooling performance. In order to quantify the attainable temperature reduction, an electro-thermal circuit model is presented, in which the key electronic and thermal transport parameters enter.  Overall, our results demonstrate that the studied OPE3 devices are compatible with temperature reductions of several K. Based on the results, some strategies to enable high performance devices for cooling applications are briefly discussed.  
\end{abstract}


\maketitle
\section{Introduction} \label{sec:intro}
The advance of experimental techniques to measure the transport properties of molecular junctions \cite{Cuevas2010,nitzan2003electron,song2011single} has motivated many theoreticians to investigate these devices \cite{koentopp2008density,evers2020advances}. Most  theoretical studies have been done considering the system at equilibrium, that is without an applied bias potential. This is motivated by the fact that key characterizing parameters of the junction like the conductance $G$, the Seebeck coefficient $S$ and the figure of merit $ZT$ for thermoelectric applications are defined as linear response properties in the limit of vanishing bias \cite{nikolic2012first}. In addition, out-of-equilibrium simulations are technically more demanding, as the electro-static potential in the molecular region needs to be  determined accurately. On the experimental side, the determination of currents under sizable bias is likewise difficult, given that the electric field can destabilize the delicate metal-molecule bonding. However, to take full advantage of a molecular junction, identifying charge and energy transport  out of equilibrium is a necessity \cite{darancet2012quantitative,tan2010measurement}. Studying the I-V characteristics of molecular junctions provides a complete picture of the transport mechanisms at play.

In this field, first principles calculations are a powerful tool to accompany and rationalize experimental data. It has been shown that the Non-equilibrium Green's function (NEGF) method in conjunction with Density Functional Theory (DFT) is well suited to calculate electronic properties of single molecular junctions with or without applied bias \cite{brandbyge2002density,darancet2012quantitative}.
\begin{figure}[htpb]
    \centering
    \includegraphics[width=0.5\textwidth]{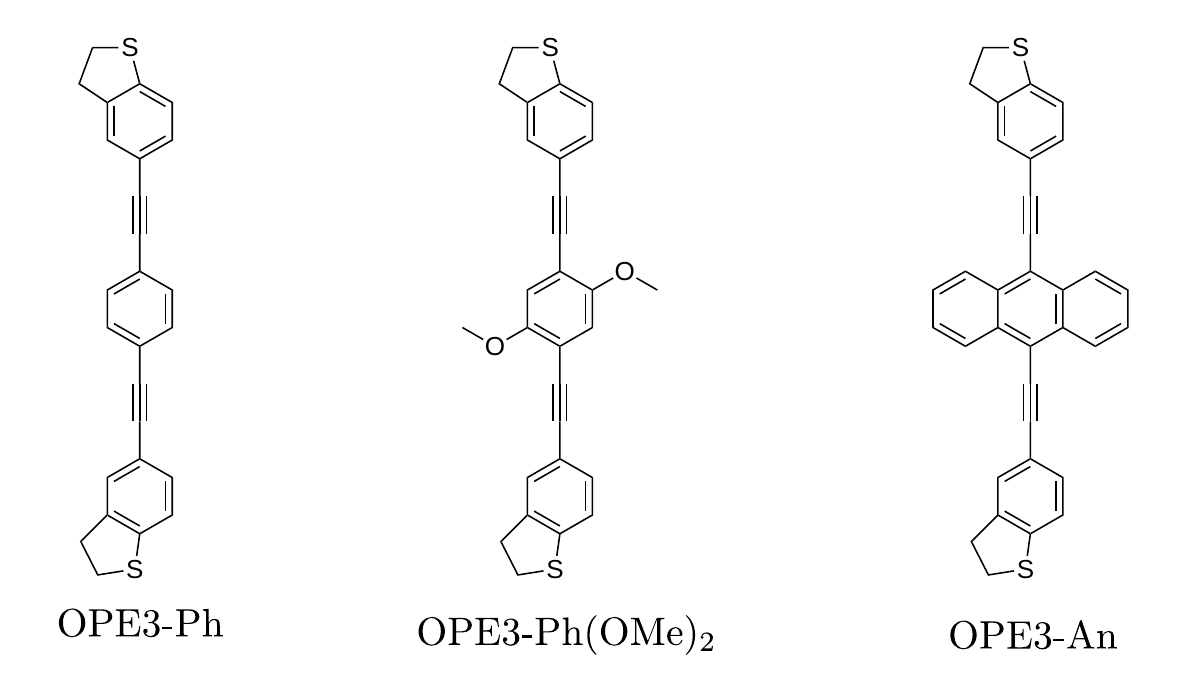}
    \caption{OPE3 derivatives with dihydrobenzo[b]thiophene (DHBT) anchoring groups and different side chains. Reproduced from \cite{dekkiche2020electronic}.}
    \label{ope3-deravative-4ch}
\end{figure}
Besides charge transport, recent efforts in the field of molecular electronics focus also on the question how heat is transported and dissipated in the device. Both electrons and phonons participate in this process, while photon based transfer can typically be neglected. Measurements of the Seebeck coefficient quantify the induced voltage by an applied temperature gradient and are important to investigate the electronic channel \cite{tan2010measurement,widawsky2012simultaneous,guo2013single}. In a recent study, direct Peltier cooling through molecular junctions was demonstrated\cite{cui2018peltier}, which is a crucial first step to develop bottom-up cooling devices \cite{ding2021advanced}.

In this study, we combine parallel electronic and phononic channels of heat transport in a electro-thermal circuit model to provide realistic estimates of the temperature decrease that can be actually reached in prototypical molecular junctions. To this end we study molecular junctions incorporating three oligo
(phenyleneethynylene) derivatives (OPE3) (Fig.~\ref{ope3-deravative-4ch}) which have been previously characterized in equilibrium both experimentally and theoretically \cite{dekkiche2020electronic,dekkiche2021correction}. In a first step, we compute the current-voltage curves using a NEGF formalism based on the approximate DFT method DFTB \cite{seifert1996calculations,elstner1998self,pecchia2008non}. The bias-dependent transmission is then employed to calculate the dissipated heat in the electrodes over a wide range of applied voltages and temperature gradients. Optimal parameter ranges are determined and used in an electro-thermal circuit model that takes into account the phonon heat backflow through the molecule  and the parasitic heat leakage of typical experimental setups. It will be shown that under these optimal conditions a cooling power of several nW through a molecule can be reached.
\section{Methods} \label{sec:method}
\begin{figure}[htbp]
    \centering
    \includegraphics[width=0.48\textwidth]{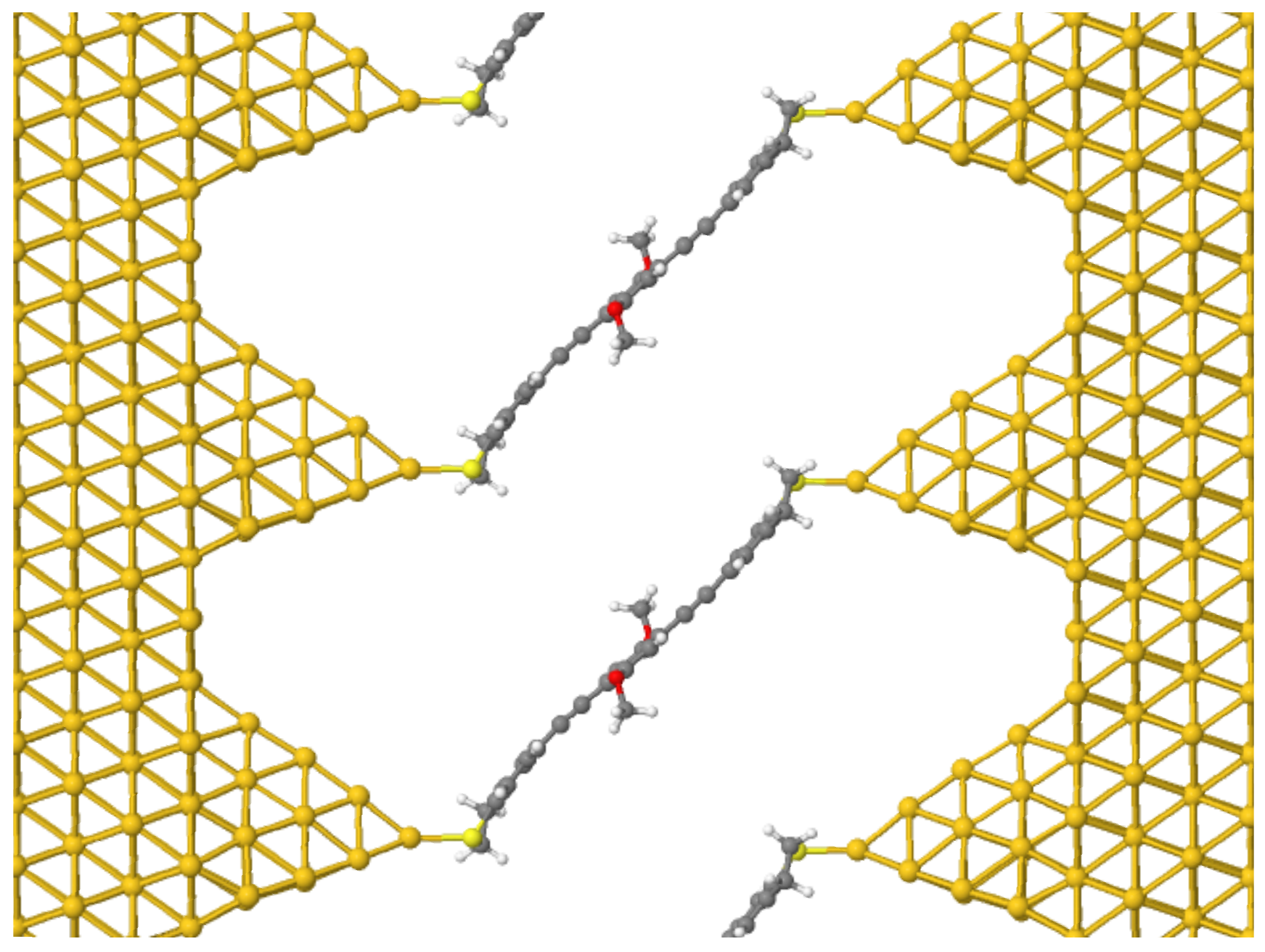}
    \caption{Device geometry for DFTB transport simulations. Shown here is
{\bf OPE3-Ph(OMe)$_2$} connected to semi-infinite Au(111) leads. The molecules feature a DHBT anchor group (see Fig.~\ref{ope3-deravative-4ch}) that binds to an Au$_{20}$ cluster, mimicking the tip of a Scanning Tunneling Microscope (STM). Reproduced from \cite{dekkiche2020electronic}.}
    \label{device}
\end{figure}

In the past, the approximate DFT method DFTB has been applied successfully to compute transport characteristics of various molecular devices \cite{wang2013thermoelectric,yam2011multiscale,shenogin2020effect,zhang2014quantum,ghorbani2013strain,dekkiche2020electronic,pecchia2008non}. DFTB is characterized by a second-order expansion of the DFT total energy functional around a suitably chosen reference density. The methods numerical efficiency stems from pre-calculated Hamiltonian matrix elements and the partial neglect and approximation of two-electron integrals \cite{seifert1996calculations,frauenheim2002atomistic}. Here we use DFTB combined with NEGF theory as implemented in the {\tt DFTB+} code \cite{hourahine2020dftb+}. As described in more detail in Ref.~\onlinecite{pecchia2008non}, the transmission $t(E,V)$ is computed as 
\begin{equation}\label{tau}
   t=\text{Tr}(\Gamma_{L}G^{r} \Gamma_{R} G^{a}), 
 \end{equation}
where $G^r$ and $G^a$ are retarded and advanced Green's functions and $\Gamma_{L/R}$ corresponds to matrices that describe the coupling of the molecule with the left (L) and right (R) lead. Note that all these terms depend on the applied bias $V$ and are computed using a self-consistent determination of the electro-static potential in the device region using a Poisson solver. The current can then be obtained from 
\begin{equation}
   I(V)= \frac{2e}{h}\int_{-\infty}^{\infty} t(E,V)\left[f_L(E)-f_R(E)\right]\, dE, 
\end{equation}
where $f_L$ and $f_R$ are the Fermi-Dirac distributions in the left and right lead, which depend on the chemical potential  $\mu_{L/R}=E_F \pm |e|V/2$. Here $E_F$ denotes the Fermi energy and the bias is applied symmetrically. The device geometry for the different OPE3 derivatives was generated by a combination of periodic and gas-phase DFT calculations as detailed in our previous publication \cite{dekkiche2020electronic} and is shown in Fig.~\ref{device}. In total the device region consists of 266 (for {\bf  OPE3-Ph}), 274 (for {\bf OPE3-Ph(OMe)$_2$}), 278 (for {\bf OPE3-An}) atoms and encompasses the molecule bound to the Au(111) surface by Au$_{20}$ clusters and three additional layers of bulk Au. The semi-infinite leads are modelled by six Au layers each. We used the auorg-1-1 Slater-Koster set \cite{elstner1998self,niehaus2001application,fihey2015scc}  with orbital dependent Hubbard parameters for Hamiltonian and overlap construction. For the NEGF-DFTB calculations, a real-space Poisson solver using periodic boundary conditions was employed to calculate the potential and density matrix in the device region. The default values for the Poisson and Green's function solver were used \cite{hourahine2020dftb+}. Perpendicular to the transport direction the Brillouin zone was sampled by a 10$\times$10 Monkhorst-Pack set.

\section{Results and discussions} \label{sec:result}
\subsection{Current-Voltage characteristics}
\begin{figure}[htpb]
    \centering
    \includegraphics[width=0.5\textwidth]{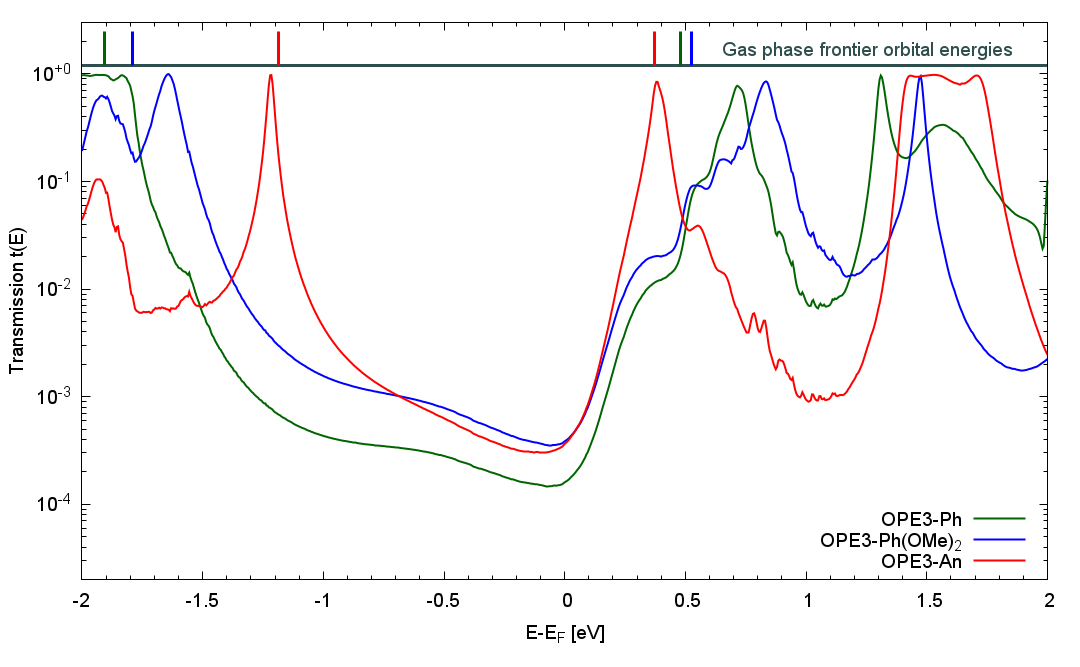}
    \caption{DFTB transmission for OPE3 derivatives as investigated in \cite{dekkiche2020electronic}. For
illustrative purposes the DFTB frontier orbital energies of the isolated
molecules in the gas phase are given at the top of the figure. Reproduced from \cite{dekkiche2021correction}.}
    \label{trans0}
\end{figure}
Before discussing the bias dependence of the electronic transport, it seems worthwhile to review the equilibrium properties of the three OPE3 derivatives shortly. Fig~\ref{trans0} shows the transmission function $t(E)$ taken from Ref.~\onlinecite{dekkiche2021correction}. The key finding is that all molecules feature a lowest unoccupied molecular orbital (LUMO) that is closer to the Fermi energy than the highest occupied molecular orbital (HOMO). This indicates that for weak bias transport occurs predominantly through the tails of the LUMO resonance. The Seebeck coefficient (also called thermopower) 
\begin{equation}
    \label{See} S = - \frac{\pi^2k_B^2 T}{3e} \frac{t'(E_F)}{t(E_F)},
\end{equation}
is hence negative and was found to be -25.6 $\mu$V/K for {\bf OPE3-Ph}, -27.0 $\mu$V/K for {\bf OPE3-(OMe)$_2$} and -37.1  $\mu$V/K for {\bf OPE3-An} in NEGF-DFTB simulations \cite{dekkiche2021correction}.  In the same study, the experiments confirmed the main transport mechanism, though for the thermopower more positive values were obtained. The functional groups in {\bf OPE3-Ph(OMe)$_2$} and {\bf OPE3-An} affect mainly the HOMO level, while the  energetical position of the LUMO resonance is the same, such that rather small variations of $S$ are found.

\begin{figure*}
    \centering
    \includegraphics[width=\textwidth]{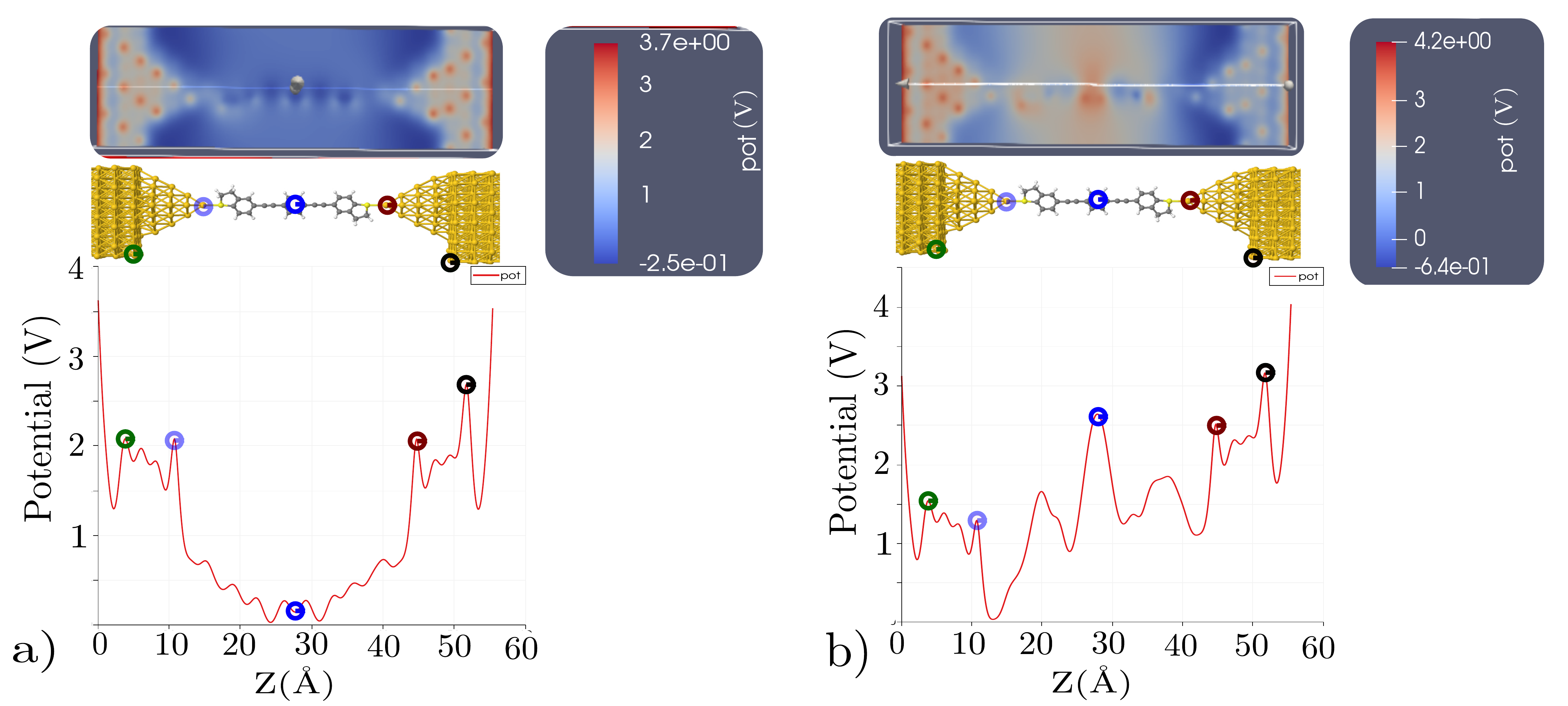}
    \caption{(Top panel) 2D color map of the electrostatic potential ($xy$-plane crossing the apexes of the gold pyramids)  in the device region for {\bf  OPE3-Ph} and sketch of the device region. The scale bar on the top right is in units of V. (bottom panel) Potential along the line indicated in the top panel. Small circles with the same color show the same coordinate in transport direction. \label{poten} a) {\bf OPE3-Ph} under bias 0 V, b) {\bf OPE3-Ph} under bias -1 V}
    
\end{figure*}
We now go beyond the investigations carried out in Refs. \onlinecite{dekkiche2020electronic,dekkiche2021correction}. We apply a  finite symmetric bias voltage to both leads, such that the chemical potential of the left and the right lead changes to $\mu_R=E_F-eV/2$ and $\mu_L=E_F+eV/2$, respectively, where $e=|e|$ denotes the absolute value of the electron charge. Hence, a current flows from the left lead to the right lead for negative bias. Non-equilibrium transport simulations require a self-consistent cycle in which the Hamiltonian is iteratively updated with the device potential, which in turn leads to a change in the charge density used in the Poisson equation. For large bias values the electronic structure is strongly perturbed which may lead to convergence difficulties. In our simulations we were able to obtain results for bias values up to $\pm 1$ V.

In Fig.~\ref{poten} we show exemplarily the potential in the device region for {\bf OPE3-Ph} coupled perpendicularly to the leads at $V=-1$ V and $V=0$ V.  At equilibrium, the potential is nearly symmetrical with respect to the center of the molecule, which is expected since the device is symmetrical.  For the junction under bias, the potential drops linearly along the molecular part with some fluctuations on the atomic scale. The potential in the metallic Au$_{20}$ pyramids is similar to the equilibrium and just shifted by $\pm$ 0.5 V as a whole, indicating a rather efficient screening of the field in this small cluster.

\begin{figure*}
    \centering
    \includegraphics{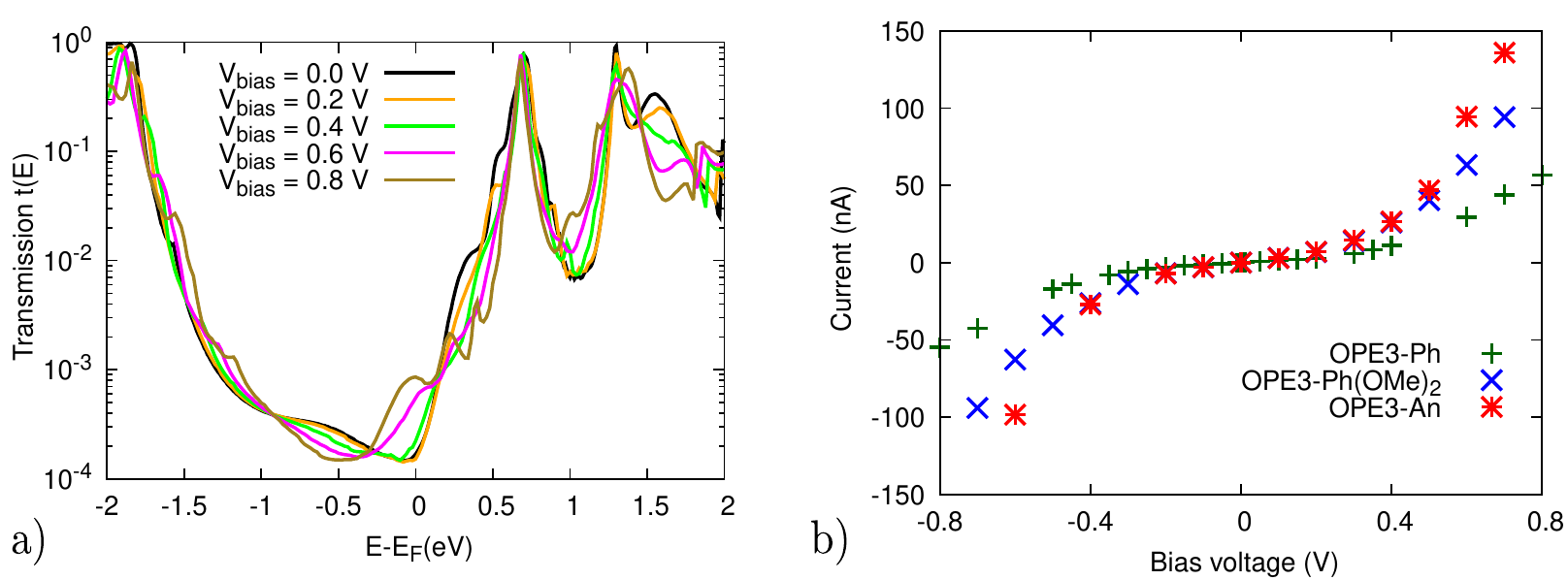}
    \caption{a) Evolution of the electronic transmission under bias for {\bf OPE3-Ph}, b) The current as a function of bias for OPE3 derivatives.}
    \label{trans-evol}
\end{figure*}

The applied bias has a non-neglible effect on the transmission of the junction as shown in Fig.~\ref{trans-evol}a exemplarily for {\bf OPE3-Ph}. For $V$ in the range [0:0.8] V, the LUMO resonance (around 0.4 eV) does not shift strongly in energy but exhibits a reduced transmission at higher bias. Large changes are also found around $E-E_F=-0.5$ V, where the weakly transmitting resonance diminishes as the bias increases and a new feature directly at the Fermi energy arises. In Ref.~\onlinecite{Taba2022}, the corresponding state at $E-E_F=-0.5$ V was tentatively assigned to a localized state at the molecule-Au$_{20}$ interface, which has no gas phase counterpart in contrast to the HOMO/LUMO frontier orbitals that keep their spatial form also in the metal-molecule-metal complex.

The current-voltage  curves for the different OPE3 derivatives are shown in Fig.~\ref{trans-evol}b. In general, the I-V characteristics  are nonlinear, as reported for other organic molecular junctions \cite{lee2013heat,selzer2004thermally,lee2013heat}. Additionally, the symmetry of I-V curves proves that the coupling to the left and right leads is very similar in our transport device \cite{selzer2004thermally,perrin2020single}. As the bias value increases, {\bf OPE3-An} and {\bf OPE3-Ph(OMe)$_2$} show a higher value of current compared to {\bf OPE3-Ph}. This is due to the larger transmission around the Fermi energy (c.f.~Fig.~\ref{trans0}) for these two molecules. Around $V=0.5$ V all systems show a steeper increase of the current, since the LUMO resonance centered at $\approx 0.4$ eV starts to enter the bias window which extends from $-V/2$ to $V/2$. Given the larger transmission of {\bf OPE3-An} at the LUMO resonance, the highest current values are obtained for this compound.


\subsection{Heat dissipation out of equilibrium}

\begin{figure*}
    \centering
    \includegraphics[height=0.6\textheight]{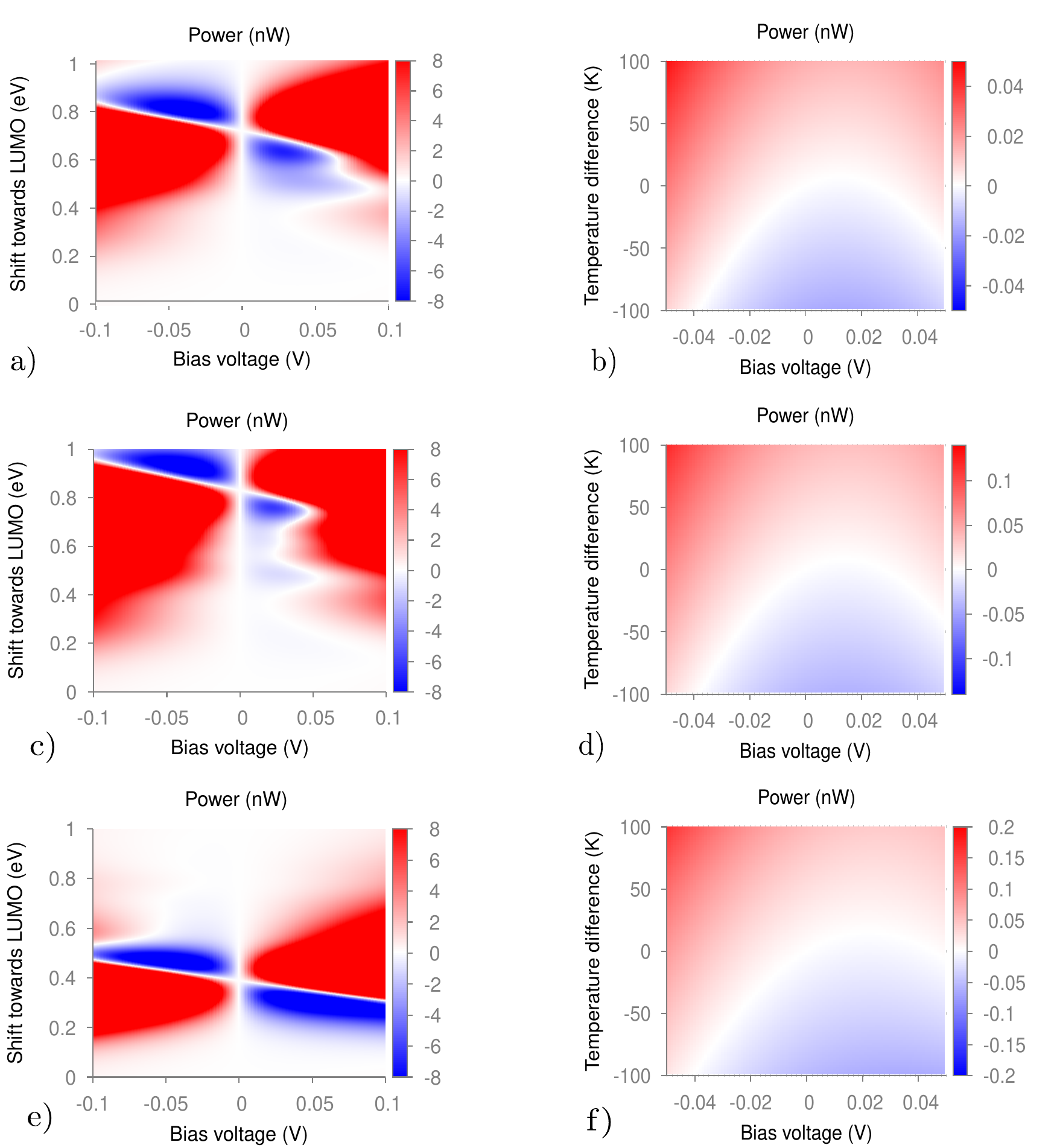}
    \caption{(left column) Bias dependent net heating/cooling in OPE3 derivatives. Shown is the power dissipated in the left lead $P_L$ for $T_L$ = $T_R = 300$ K as a function of bias voltage and the shift of the Fermi energy ($\Delta$) towards the LUMO. Red color indicates positive values (heating) and blue color  negative values (cooling). a) {\bf OPE3-Ph}, c) {\bf OPE3-Ph(OMe)$_2$}, e) {\bf OPE3-An} (right column)  $P_L$ as function of bias and $\Delta T = T_L -T_R$, with $T_R = 300$ K and $\Delta = 0$ eV. b) {\bf OPE3-Ph}, d) {\bf OPE3-Ph(OMe)$_2$}, f) {\bf OPE3-An}.\label{pl-fig}}
\end{figure*}
\begin{figure*}
\includegraphics[width=\textwidth]{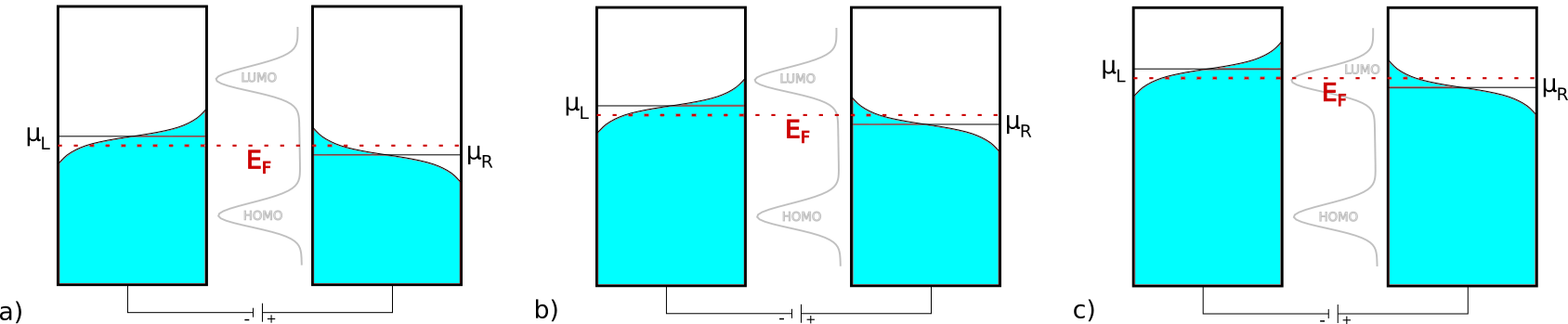}
\caption{Schematic representation of Peltier cooling in molecular devices. The right lead is at lower chemical potential (V > 0), electrons move from left to right. The occupation probability of the electrons in the two leads ($T>0$) is shown in blue color versus energy and the dotted line corresponds to the Fermi energy of the device. Gating leads to a shift of the molecular density of states, shown in grey color, which is represented here by HOMO and LUMO levels.  a) $\mu_L$ far away from LUMO, moderate cooling of the left lead, b) $\mu_L$ close to resonance, strong cooling of left lead, c)  $\mu_L$ above resonance, heating of left lead.}
\label{junc-schem}
\end{figure*}

Next, we used the bias dependent transmission to estimate the dissipated heat in the leads. The power $P_L$ associated with the total heat dissipated in the left lead is given within the Landauer-B{\"u}ttiker formalism as \cite{cui2018peltier}:
\begin{equation}\label{pl}
    \begin{aligned}
    &P_L(T_L,T_R,V)=&\\
     &\frac{2}{h}\int_{-\infty}^{\infty}(\mu_L-E) t(E,V) \left[f_L(T_L)-f_R(T_R)\right] dE.&\\ 
    \end{aligned}
\end{equation}
Note that the total dissipated heat includes both Joule heating and the Peltier effect, consequently, net refrigeration happens when P$_L < $ 0. Fig.~\ref{pl-fig} depicts the results for the different OPE3 derivatives. Shown in the first column is a 2D plot of $P_L$ as a function of bias $V$ and the shift $\Delta$ of the Fermi energy towards more positive values. Experimentally such a shift can be realized by gating in a three-terminal device \cite{song2009observation} or by means of electro-chemical gating \cite{zhang2020improving}. Our motivation to investigate the effect of modifying the Fermi energy (or equivalently, the energetical position of the molecular levels) is twofold. First, as will we explained below, the attainable cooling power increases strongly when the Fermi energy approaches a molecular resonance. Second, like for most DFT based methods, the calculation of the precise Fermi level alignment between molecule and metal is quite delicate. Our approach predicted LUMO based transport in full agreement with the experimental results, but the thermopower was overestimated in absolute terms, which speaks for a slight underestimation of the $E_\text{LUMO} - E_F$ difference. By modifying $E_F$ we arrive at more general conclusions. 

If we consider first the results for {\bf  OPE3-An } in Fig.~\ref{pl-fig}e, three different transport scenarios may be distinguished that are schematically presented in Fig.~\ref{junc-schem}. For $\Delta < 0.2$ eV (corresponding to Fig.~\ref{junc-schem}a, the Fermi level is relatively far away from the LUMO. For $V>0$, hot charge carriers in the left lead may only transfer to the right through the tails of the LUMO resonance. The factor $\mu_L-E$ in Eq.~\ref{pl} is negative in this case, but the cooling effect in the left lead is weak.  For $0.2$ eV $< \Delta < 0.4$ eV (corresponding to Fig.~\ref{junc-schem}b), hot electrons from the left lead are close to the LUMO resonance and have a high probability for transmission. The resulting current leads to cooling of the left lead. Note that this Peltier cooling is compensated by a heating of the right lead such that total dissipated heat is always given by $P_L+P_R = IV$, which corresponds to Joule heating  \cite{cui2018peltier}. Finally, for $\Delta > 0.4$ eV, the LUMO resonance is placed in the bias window $[\mu_L,\mu_R]$  or below and electrons with energy $E< \mu_L$ are predominantly transmitted. This results in heating of the left lead. Changing the bias polarity inverts the roles of left and right lead as can be seen in Fig.~\ref{pl-fig}e. From the foregoing discussion it might appear beneficial to increase the bias potential, instead of - or in addition to - gating the device, in order to achieve maximal cooling. Also in this case the chemical potential $\mu_L$ approaches the LUMO resonance and electrons transmitted at $E>\mu_L$ effectively shuffle heat from $L$ to $R$. This is however more than offset by the transport in the increased bias window which always leads to (Joule) heating. Consequently, in Fig.~\ref{pl-fig}a/c/e, cooling is only observed in a narrow region around vanishing bias.

Comparing the different OPE3 derivatives, we see that the dissipated power reflects the different transmission characteristics shown in Fig.~\ref{trans0}. {\bf OPE3-An} features ideal LUMO transmission ($t=1$) at $E-E_F=0.4$ eV which translates into the simple power spectrum Fig.~\ref{pl-fig}e discussed above. Both  {\bf OPE3-Ph} and {\bf OPE3-(OMe)$_2$} exhibit weaker LUMO transmission (t $\approx 0.01$ at $E-E_F=0.4$ eV) and have near perfect transmission only at higher energies, namely at $0.7$ V  and  $0.8$ V, respectively. From this, the broader region of cooling ($0.4$ eV $<\Delta < 0.8$ eV) observed in  Fig.~\ref{pl-fig}a/c can be understood. Interpolation of the bias dependent transmission allows us to quantify the maximum cooling power ($P_L^\text{opt}$) for each of the OPE3 derivatives used in this study. The results are presented in Tab.~\ref{pl-table} and show that nW cooling may be reached at moderate bias and gating for {\bf OPE3-An}. 
\begin{table}[htbp]
    \centering
    \begin{tabular}{c  c c c}
         Molecule& $V^\text{opt}$ (V) & $\Delta^\text{opt}$ (eV)  &  $P_L^\text{opt}$ (nW)  \\
         \hline
        {\bf  OPE3-Ph}&-0.04&0.8&-8.25\\
        {\bf  OPE3-Ph(OMe)$_2$}&-0.03&0.9&-9.29\\
        {\bf OPE3-An}&0.06&0.3&-11.29\\
         \hline
    \end{tabular}
    \caption{Maximum cooling power ($P_L^\text{opt}$) reached for optimal bias potential ($V^\text{opt}$) and Fermi energy shift ($\Delta^\text{opt}$) for $T_L = T_R = 300$ K.}
    \label{pl-table}
\end{table}

We also shortly discuss the dependence of $P_L$ on the  temperature difference ($\Delta T= T_L-T_R$) between the two leads, keeping $T_R$ fixed at $300$ K. Figures ~\ref{pl-fig}b, \ref{pl-fig}d, and \ref{pl-fig}f indicate a rather weak dependence on this parameter. At zero bias, cooling occurs if the left lead is at higher temperature as the left lead, in line with the schematic representations given in Fig.~\ref{junc-schem}a.
 
\subsection{Electro-thermal circuit for OPE3 derivatives}
In this section we provide estimates for the temperature reduction that can be reached in real devices based on a simple electro-thermal circuit. To do so, we base our discussion on the experimental setup which was used to measure the charge and heat transport for one of the molecular junctions ({\bf OPE3-Ph}) used in this study \cite{mosso2019thermal,mykkanen2020thermionic}. In the experiment, a micro-electro-mechanical system (MEMS) was suspended using four silicon nitride beams. The characteristic thermal conductance of this support can be reduced to below $\kappa^{Ph}_{supp}=$ 10$^{-8}$ W/K. In the central membrane of the MEMS device, a thermometer and a gold surface are located on which the molecules are deposited.  Heat and electronic transport measurements are then carried out upon contact with a scanning tunnelling microscope (STM) tip.

 \begin{figure}
 \includegraphics[width=0.48\textwidth]{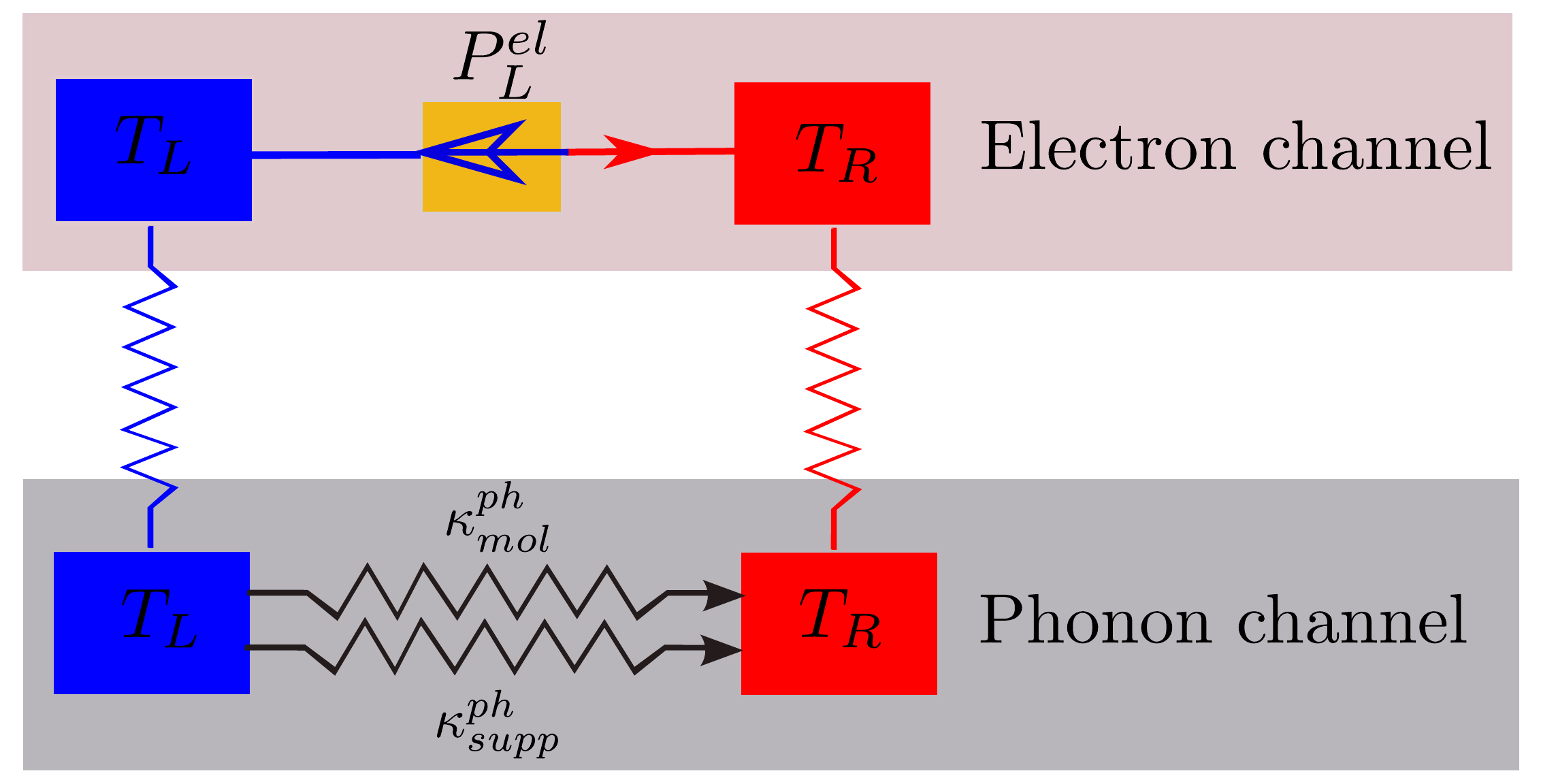} 
\caption{ Illustration of the electro-thermal circuit used to model the temperature gradient in molecular junctions.\label{circuit}}
\end{figure}

\begin{figure}
    \centering
    \includegraphics[width=0.48\textwidth]{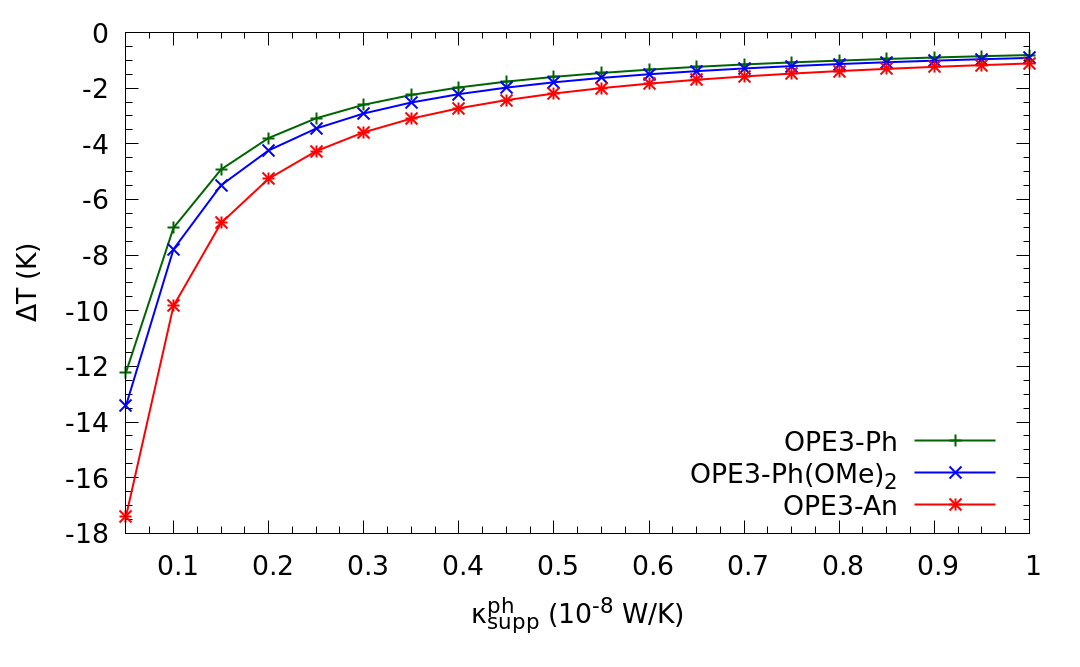}
    \caption{Temperature difference between both leads $\Delta$T versus thermal conductance of the support (T$_R$ = 300 K, $\kappa^{ph}_{mol} =  22$ pW/K). }
    \label{dt-fig}
\end{figure}

We model this setup with the electro-thermal circuit as shown in Fig.~\ref{circuit}, which was inspired by a similar model for a semiconductor-superconductor thermionic junction \cite{mykkanen2020thermionic}, where the quasiparticle gap gives similar energy filtering as the HOMO-LUMO gap in this work. Here the left lead is identified with the gold surface on the central MEMS platform, while the right lead corresponds to the STM tip. The temperature of the tip stays constant due to the good thermal contact with the environment.  When the net heating power is negative, the left lead cools down due to the Peltier effect which is of electronic origin. In the phonon channel, the thermal conductance of both molecule and the support need to be considered. 
At equilibrium, one has: 
\begin{equation}
 P_L^{el}+ P^{ph}_{supp} + P^{ph}_{mol} = 0,
 \label{pel}
\end{equation}
where $P_L^{el}<0$ is the net heating power due to the electrons in the left lead, $P^{ph}_{supp}= -\kappa^{ph}_{supp} \Delta T$ is the heat loss due to the support, and $P^{ph}_{mol}= -\kappa^{ph}_{mol} \Delta T$ gives the backflow of heat along the molecule due to phonons. The thermal conductance of {\bf OPE3-Ph} was measured to be  $22$ pW/K at room temperature \cite{Gemma2021}, while conductances for the other two OPE3 derivatives have not yet been reported. Given the fact that all studied molecules in this study share the same anchor group, we expect modest variations in their thermal conductivity. In this context we mention the study by Klöckner and co-workers \cite{klockner2017tuning} showing that quantum interference may have an important impact on the thermal conduction in molecular junctions \cite{markussen2013phonon} and can be tuned by appropriate functional groups attached to the molecular backbone. The authors study also OPE3 junctions and find variations in $\kappa^{ph}_{mol}$ of roughly 20 \%  between differently functionalized OPE3 molecules. Since $\kappa^{ph}_{mol} \ll \kappa^{ph}_{supp}$, such a variation can be neglected and we will work with the same  $\kappa^{ph}_{mol}$ for all molecules. As additional approximation we will assume that the thermal conductances are not temperature dependent. This is well justified, because the temperatures considered here are well above the Debye temperature of the gold contacts. We do, however, account for the temperature dependence of $P_L$ according to Eq.~\ref{pl} and solve Eq.~\ref{pel} in the form 
\begin{equation}
        \Delta T =\frac{P_L^{el}(\Delta T)}{\kappa^{Ph}_{supp} + \kappa^{Ph}_{mol} },
    \end{equation}
self-consistently for $\Delta T = T_L -T_R$, with $T_R = 300$ K. Due to the rather weak temperature dependence of $P_L$ as seen in Figures ~\ref{pl-fig}b, \ref{pl-fig}d, and \ref{pl-fig}f, only few iterations are typically necessary in this process. Without gating, i.e.~for $\Delta = 0$, results for all studied molecules show only very marginal cooling with $\Delta T \approx 1$ mK. Refrigeration can be significantly enhanced by gating. Taking the optimal values of Tab.~\ref{pl-table} as input parameters, the results in Fig.~\ref{dt-fig} are obtained. Here we vary also the conductance of the support to see which temperature gradients could be reached if the heat loss to the support could be further lowered by mechanical engineering. The observed temperature difference of several K for all studied molecules presents certainly an upper bound, but clearly demonstrates the potential of bottom-up architectures for cooling devices. In this context we also comment on the benefits of cross-linking the OPE3 molecules to form thin films. In such a case not only the stability of the device would be improved but also the total cooling power. Assuming that $N$ molecules act in parallel, both $P_L$ and $\kappa^{ph}_{mol}$ would increase by this factor. Since the molecular conductance would be still much smaller than the parasitic heat leakage, a nearly linear raise of the temperature gradient could be obtained.     
\section{Conclusion} \label{sec:conc}
In summary, we performed NEGF transport simulations at the DFTB level for three OPE3 derivatives that were recently synthesized and experimentally characterized. Full I-V curves were computed and show a strong current increase at around $0.4$ eV when the LUMO resonance enters the window of conduction. {\bf OPE3-Au} exhibits the largest currents in the studied bias range, which can be explained by its enhanced LUMO transmission compared to the other two molecules.  The voltage bias dependent transmission was then used to quantify  the heat transport through the molecules under non-equilibrium conditions. It was found that junction gating has a profound beneficial impact on the cooling power which can reach several nW under optimal conditions. Cooling is observed at small bias voltages and large bias leads to simple Ohmic heating. Given that good electrostatic control of molecular junctions is still difficult to achieve, a viable strategy might consist of pushing the frontier orbitals closer to the Fermi energy. This could be achieved by appropriate electron-withdrawing or donating functional groups or using quantum interference effects as recently discussed by several groups \cite{sadeghi2019quantum,jia2018quantum}. Finally, we used a combination of experimental data and the theoretical results of this study to set up an electro-thermal circuit model that combines the electronic and phononic transport channels. As a result, we obtained promising values  for the attainable temperature reduction of several K in these bottom-up molecular devices. Further improvements are expected for cross-linked molecular architectures. Simulations of these more complex devices are currently under way.      

\section*{Acknowledgements} \label{sec:acknowledgements}
We acknowledge the European Commission H2020 projects ‘EFINED’ http://www.efined-h2020.eu Grant Agreement no.766853  for providing financial resources to our project. We thank the GENCI for computational resources (under project DARI A0050810637 and A0070810637). MP acknowledges financial support of the Academy of Finland through project HyPhEN (No 342586) and Centre of Excellence program No 336817. We thank the EFINED team for inspiring discussions.    
  
\bibliography{bib}


\end{document}

%% file: preamble.tex
\usepackage{amsthm}
\usepackage{mathtools}
\usepackage{physics}
\usepackage{xcolor}
\usepackage{graphicx}
\usepackage[left=23mm,right=13mm,top=35mm,columnsep=15pt]{geometry} 
\usepackage{adjustbox}
\usepackage{placeins}
\usepackage[T1]{fontenc}
\usepackage{lipsum}
\usepackage{csquotes}

\usepackage{amsmath}